\documentclass[runinaddress,showpacs,twocolumn,prb]{revtex4}%
\usepackage{amssymb}
\usepackage{amsmath}
\usepackage{graphicx}
\usepackage{epstopdf}
\usepackage{amsfonts}

\begin{document}
\title{Disorder-induced topological phase transitions on Lieb lattices}
\date{\today }
\author{Rui Chen}
\author{Dong-Hui Xu\footnote{donghuixu@hubu.edu.cn}}
\author{Bin Zhou\footnote{binzhou@hubu.edu.cn}}
\affiliation{Department of Physics, Hubei University, Wuhan 430062, China}

\begin{abstract}
	
Motivated by the very recent experimental realization of electronic Lieb lattices and research interest on topological states of matter, we study the topological phase transitions driven by Anderson-type disorder on spin-orbit coupled Lieb lattices in the presence of spin-independent and dependent staggered potentials. By combining the recursive Green's function and self-consistent Born approximation methods, we found that both time-reversal-invariant and time-reversal-symmetry-broken spin-orbit coupled Lieb lattice systems can host the disorder-induced gapful topological phases, including the quantum spin Hall insulator (QSHI) and quantum anomalous Hall insulator (QAHI) phases. For the time-reversal-invariant case, the disorder induces a topological phase transition directly from a normal insulator (NI) to the QSHI. While for the time-reversal-symmetry-broken case, the disorder can induce either a QAHI-QSHI phase transition or a NI-QAHI-QSHI phase transition, depending on the initial state of the system. Remarkably, the time-reversal-symmetry-broken QSHI phase can be induced by Anderson-type disorder on the spin-orbit coupled Lieb lattices without time-reversal symmetry.
\end{abstract}

\maketitle

\section{Introduction}

The Lieb lattice \cite{Lieb89PRL} is a line-centered square lattice, characterized by the unit cell containing three sites with unequal connectivity. This special geometric structure results in a peculiar band spectrum consisting of a flat band with zero energy touching two linearly dispersive intersecting bands at a single Dirac point. The Lieb lattice may serve as an ideal platform to study quantum many-body physics, such as ferromagnetism \cite{Lieb89PRL,Tasaki99PTP,Mielke91JPAM,SSQ94PRL} and high temperature superconductivity \cite{Miyahara07PhyC,Julku16PRL,Kopnin11PRB}, due to its special band structure. The Lieb lattice has been extensively investigated both theoretically and experimentally for its unique band structure in recent years. Although the Lieb lattices have been realized in photonic crystals \cite{Vicenio15PRL,Mukherjee15PRL} and cold atom systems \cite{Shen10PRB,Apaja10PRA,Goldman11PRA}, creating an electronic version has proved more challenging. Recently, Qiu \emph{et al}. showed that the artificial Lieb lattices can be realized on the metallic copper surface \cite{Qiu16PRB}. The artificial Lieb Lattice was rapidly confirmed in experiments \cite {Drost17NatPhys,Slot17NatPhys}, providing a realistic electronic Lieb lattice system to explore the above-mentioned physics.

Nowadays, the search for topological phases of matter has been a fundamental theme of condensed matter physics. It was proposed that the QSHI phase and the QAHI phase can be realized on the Lieb lattices with intrinsic spin-orbit coupling \cite{Weeks12PRB,Weeks10PRB,Goldman11PRA,Zhao12PRB,Beugeling12PRB,Tsai15NJP,Nita13PRB,Chen17PLA}. Moreover, the QSHI phase on the artificial honeycomb lattice was proposed to be realized by depositing molecules on the surface of heavy metals, such as gold which has a strongly spin-orbit coupled surface \cite{Hughes12PRB}. We believe the realization of the QSHI phase on the artificial Lieb lattice is also promising by using a heavy metal as the substrate in the current artificial systems .

It is thought that the QSHI phase is robust against weak disorder due to the non-trivial topology of bulk state \cite{Kane05PRL1,Kane05PRL2,Sheng05PRL,Taskin12PRL}, and it collapses under strong disorder. Interestingly, recent years have seen that moderate disorder can convert a topologically trivial phase to a topologically non-trivial phase. The topological Anderson insulator (TAI) is one such disorder-induced topological phase \cite{LiJ09PRL,Jiang09PRB,Groth09PRL,Jiang16cpb}. The TAI has been investigated in many related models and systems, such as Haldane model, Kane-Mele model, the three dimensional Dirac-Wilson model and semimetal systems \cite{Xing11PRB,Orth16Scirep,GUO10PRL,Guo11PRB,Guo10PRB,Chen17PRB}. Topological phase transitions driven by disorder and transport properties have been studied in various electronic systems, in which many intriguing phenomena were observed due to the interplay between topology and disorder \cite{Su16PRB,Liu16PRL,Qin15arxiv,Kimme16PRB,ChenCz15PRL,Su17arxiv,Shapourian16PRB,ChenCZ15PRB}.

In this paper, we investigate disorder effects on a spin-orbit coupled Lieb lattice model with spin-independent and dependent staggered potentials by use of combining the numerical simulation method based on the recursive Green's function and the self-consistent Born approximation. When only the spin-independent staggered potential exists, the system respects the time-reversal symmetry, we found that disorder can drive a topologically trivial phase to the QSHI phase. This is similar to the case of the Bernevig-Hughes-Zhang (BHZ) model, in which the topological Anderson insulator was revealed. When we turn on both the spin-independent and the spin-dependent potentials, the time-reversal symmetry is broken, we found that Anderson disorder can cause either a QAHI-QSHI phase transition or a NI-QAHI-QSHI phase transition depending on the initial state of the system. Different from the case of the BHZ model, disorder can induce both the QAHI phase and the time-reversal-symmetry-broken QSHI phase in our case. More interestingly, we can switch the QAHI and the QSHI phases by tuning the
strength of disorder.

\section{Model}

\begin{figure}[ptb]
\includegraphics[width=8cm]{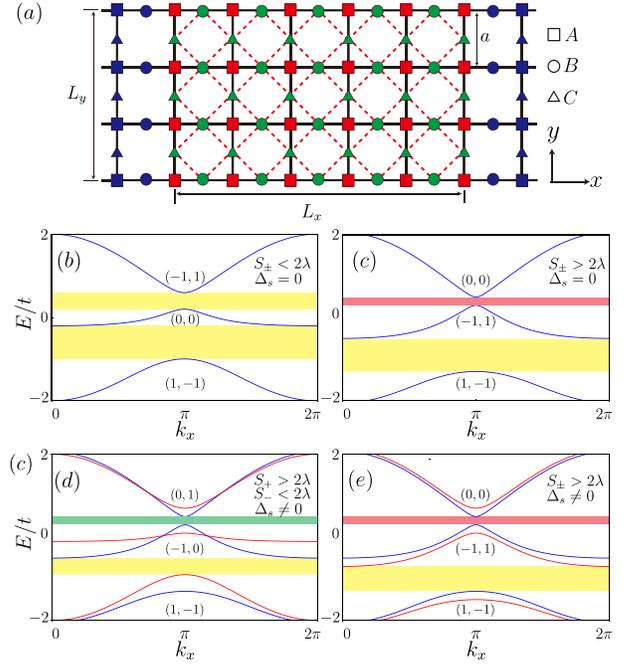} \caption{(Color online) (a) Schematic illustration of the Lieb lattice ribbon used in our transport simulations. The unit cell of Lieb lattice consists of three sublattices $A$ (the square dot), $B$ (the circle dot) and $C$ (the triangle dot). The two metal leads (blue site regions) are modeled by the tight-binding model Hamiltonian with only NN hopping on Lieb lattices. Different colors of sublattice represent distinct staggered potentials. (b-e) The bulk spectrum of system for $k_{y}=\pi$. The parameters are given by (b) $S_\pm<2\lambda$, $\Delta_{s}=0$, (c) $S_\pm>2\lambda$, $\Delta_{s}=0$, (d) $S_+>2\lambda$, $S_-<2\lambda$, $\Delta_{s}\neq0$, and (e) $S_\pm>2\lambda$, $\Delta_{s}\neq0$. The blue and red lines correspond to the spin-up and down components of the bulk band, respectively. The energy gaps labeled with red, yellow and green correspond to the topologically trivial, quantum spin Hall and quantum anomalous Hall gaps, respectively. The spin Chern number of each band is labeled in the form of $(C^{s}_+,C^{s}_-)$ with $C^{s}_{\pm}=0,\pm 1$. We plot only the spin-up component of the bulk band in (b) and (c) since the two spin components are degenerate.}%
\label{fig1}%
\end{figure}

We start with the tight-binding model with intrinsic spin-orbit coupling on the Lieb lattice. The sublattices of the Lieb lattice are labeled $A$, $B$ and $C$ as shown in Fig. \ref{fig1}(a). Then the model Hamiltonian we consider is expressed as
\begin{equation}
\mathcal{H}=\mathcal{H}_{0}+\mathcal{H}_{I}+\mathcal{H}_{c}+\mathcal{H}%
_{s}\text{.} \label{H}%
\end{equation}
The first term of the Hamiltonian (\ref{H}) reads%
\begin{equation}
\mathcal{H}_{0}=t\sum_{\left\langle ij\right\rangle, \sigma }c_{i,\sigma}^{\dagger}%
c_{j,\sigma}-\mu\sum_{i,\sigma}c^{\dagger}_{i,\sigma}c_{i,\sigma}\text{,}%
\end{equation}
where $t$ is the nearest-neighbor (NN) hopping integral and $\mu$ the chemical potential,
 $c_{i,\sigma}(c_{i,\sigma }^{\dagger})$ is the annihilation
(creation) operator of electron with spin $\sigma=\uparrow,  \downarrow $ on site $i$. The second term of $\mathcal{H}$ is the intrinsic spin-orbit coupling (ISOC) term which is expressed as
\begin{equation}
\mathcal{H}_{I}=i\lambda\sum_{\left\langle \left\langle ij\right\rangle
\right\rangle,\sigma,\sigma' }\nu_{ij}c_{i,\sigma}^{\dagger}(\sigma_{z})_{\sigma\sigma'}c_{j,\sigma'}\text{,}%
\end{equation}
which corresponds to the spin-dependent next-nearest-neighbor (NNN) hopping
(as shown by dashed lines in Fig. 1(a)) with the amplitude of $\lambda$.
$\nu_{ij}=(\mathbf{d}_{ij}^{1}\times\mathbf{d}_{ij}^{2})=\pm1$, where
$\mathbf{d}_{ij}^{1}$ and $\mathbf{d}_{ij}^{2}$ are two unit vectors along
the NNN bonds connecting site $i$ to its NNN site $j$.
The spin-independent staggered potential term $\mathcal{H}_c$ is given by%
\begin{equation}
\mathcal{H}_{c}=\sum_{i\in A,\sigma}\Delta_{c}c_{i,\sigma}^{\dagger}c_{i,\sigma}-%
\sum_{i\in B,C,\sigma}\Delta_{c}c_{i,\sigma}^{\dagger}c_{i,\sigma}\text{,}%
\end{equation}
and the spin-dependent staggered potential term $\mathcal{H}_c$ is
\begin{equation}
\mathcal{H}_{s}=\sum_{i\in A,\sigma,\sigma'}\Delta_{s}c_{i,\sigma}^{\dagger}(\sigma_{z})_{\sigma\sigma'}c_{i,\sigma'}%
-\sum_{i\in B,C,\sigma,\sigma'}\Delta_{s}c_{i,\sigma}^{\dagger}(\sigma_{z})_{\sigma\sigma'}c_{i,\sigma'}\text{,}%
\end{equation}
where $\Delta_{c}$ and $\Delta_{s}$ are the magnitudes of the spin-independent
and the spin-dependent staggered potentials respectively. Obviously, the spin-dependent staggered potential term breaks the time-reversal symmetry of the system.

After the Fourier transformation, we obtain the Hamiltonian in the momentum space
\begin{equation}
\mathcal{H}=\sum_{\mathbf{k}}\Psi_{\mathbf{k}}^{\dagger}H(\mathbf{k}%
)\Psi_{\mathbf{k}}\text{,}%
\end{equation}
where $\Psi_{\mathbf{k}}=(c_{A,\mathbf{k,}\uparrow},c_{B,\mathbf{k,}%
\uparrow},c_{C,\mathbf{k,}\uparrow},c_{A,\mathbf{k,}\downarrow}%
,c_{B,\mathbf{k,}\downarrow},c_{C,\mathbf{k,}\downarrow})^T$. Then in this basis, the Hamiltonian matrix is given by
\begin{equation}
H(\mathbf{k})=%
\begin{pmatrix}
h^{+}\left(  \mathbf{k}\right)  & 0\\
0 & h^{-}\left(  \mathbf{k}\right)
\end{pmatrix}
\text{,} \label{Mod1}%
\end{equation}
with%
\begin{equation}
h^{\pm}\left(  \mathbf{k}\right)  =%
\begin{pmatrix}
S_{\pm} & 2tC_{x} & 2tC_{y}\\
2tC_{x} & -S_{\pm} & \mp4i\lambda S_{x}S_{y}\\
2tC_{y} & \pm4i\lambda S_{x}S_{y} & -S_{\pm}%
\end{pmatrix}
\text{,}%
\end{equation}
where $S_{\pm}=\Delta_{c}\pm\Delta_{s}$, $S_{u}=\sin(k_{u}/2)$ and $C_{u}%
=\cos(k_{u}/2)$ with $u=x,y$. For simplicity, in the following calculations we set the lattice constant of the tight-binding model as $a=1$ and take the strength of ISOC as $\lambda=0.3t$, and we only consider the case of positive staggered potentials ($\Delta_c$ and $\Delta_s$) and impose the constraint $\Delta_s<\Delta_c$.

The ISOC opens two bulk gaps (the upper and lower gaps in Fig. \ref{fig1}(b)) with nontrivial topology on Lieb lattices, giving rise to the QSHI state characterized by the $Z_{2}$ topological invariant $\nu=1$ \cite{Weeks10PRB}. Meanwhile, the staggered potentials play important roles in tuning the topological phases in this system \cite{Zhao12PRB}. In this paper, we use the spin Chern number \cite{Sheng06PRL,Prodan09PRB,Li10PRB,Yang11PRL} to characterize the topological phases of the pure system. It has been shown that the spin Chern number and $Z_{2}$ topological invariant would yield the same classification by investigating the topological properties of time-reversal-invariant systems \cite{Fukui07PRB}. Moreover, the spin Chern number works for both time-reversal-invariant and time-reversal-symmetry-broken cases.

The bulk band structure and topological invariant in different parameter regions are shown in Figs. \ref{fig1}(b-e). The energy gaps labeled with red color is topologically trivial, corresponding to the total spin Chern number of the occupied bands $(C^{s}_+,C^{s}_-)=(0,0)$ with $C^{s}_+ (C^{s}_- )$ defined in the spin up (down) space. While the energy gaps labeled with yellow color have $(C^{s}_+,C^{s}_-)=(1,-1)$, indicating the QSHI state with helical edge states. The energy gaps labeled with green color own $(C^{s}_+,C^{s}_-)=(0,-1)$, which corresponds to the QAHI state with chiral edge states.

In the absence of spin-dependent staggered potential ($\Delta_{s}=0$), the bulk states are double degenerate. The system undergoes a phase transition from the QSHI phase to the NI phase as increasing the strength of the staggered potential $\Delta_{c}$. When we turn on $\Delta_{s}$, the double-spin-degenerate bands are split since the time-reversal symmetry is broken. Depending on the parameters, the upper gap may exhibit the QAHI behavior or NI behavior. However, the lower gap shows the QSHI behavior and are very robust to the variation of the staggered potentials even when the time-reversal symmetry is broken. In the following sections, we mainly concentrate on the physics when the chemical potential is inside or around the upper gap since the lower gap is boring, compared with the upper gap. Actually, the time-reversal-symmetry-broken QSHI state has been found in the Rashba spin-orbit coupled honeycomb lattices \cite{Yang11PRL}. However, different from the honeycomb lattice model, the spin-orbit coupled Lieb lattice model possesses two band gaps and they can exhibit distinct topology by tuning the staggered potentials.

\section{Numerical Simulation}

In this section, we investigate the transport properties of a two-terminal setup to identify the topological phase transitions in the presence of Anderson-type disorder by using the Landauer-B\"uttiker-Fisher-Lee formula
\cite{Landauer70PhiMag,Buttiker88PRB,Fisher81PRB} as well as the recursive Green's
function method \cite{MacKinnon85PhysBCM,Metalidis05PRB}. The two terminal
conductance can be obtained by $G=(e^{2}/h)T$, where $T=$Tr$\left[
\Gamma_{L} G^{r} \Gamma_{R} G^{a} \right] $ is the transmission coefficient,
the linewidth function $\Gamma_{\alpha}=i\left[  \Sigma_{\alpha}%
^{r}-\Sigma_{\alpha}^{a}\right]  $ with $\alpha=L,R$. The retarded and advanced Green's
functions $G^{r/a}$ are calculated from $G^{r}=\left(
G^{a}\right)  ^{\dag}=\left(  \mu I-H_{C}-\Sigma_{L}^{r}-\Sigma_{R}^{r}
\right)  ^{-1}$, where $H_{C}$ is the Hamiltonian matrix of the central
scattering region, $\Sigma_{L,R}^{(r/a)}$ are the self-energies due to the attached leads.

The setup consists of a two dimensional sample of size $L_{x}\times L_{y}$ and two
semi-infinite metal leads connected to the sample along the $x$ direction as
shown in Fig. \ref{fig1}(a). To avoid the redundant scattering from mismatched interfaces between the sample and the leads, the two metallic leads are modeled by the spin-orbit coupling free tight-binding model on Lieb lattice. In our numerical simulations, we set the chemical potentials of leads $\mu_{L,R}=1.2t$ to guarantee a high density of states. We introduce the Anderson-type disorder to the
central scattering region through random on-site energy, having a uniform distribution
within the energy interval $[-U/2,U/2]$, where $U$ is the disorder strength. This setup allows us to calculate both the longitudinal conductance $G$ and the distribution of local current.

\subsection{Spin-independent staggered potential}

\begin{figure}[ptb]
\includegraphics[width=8cm]{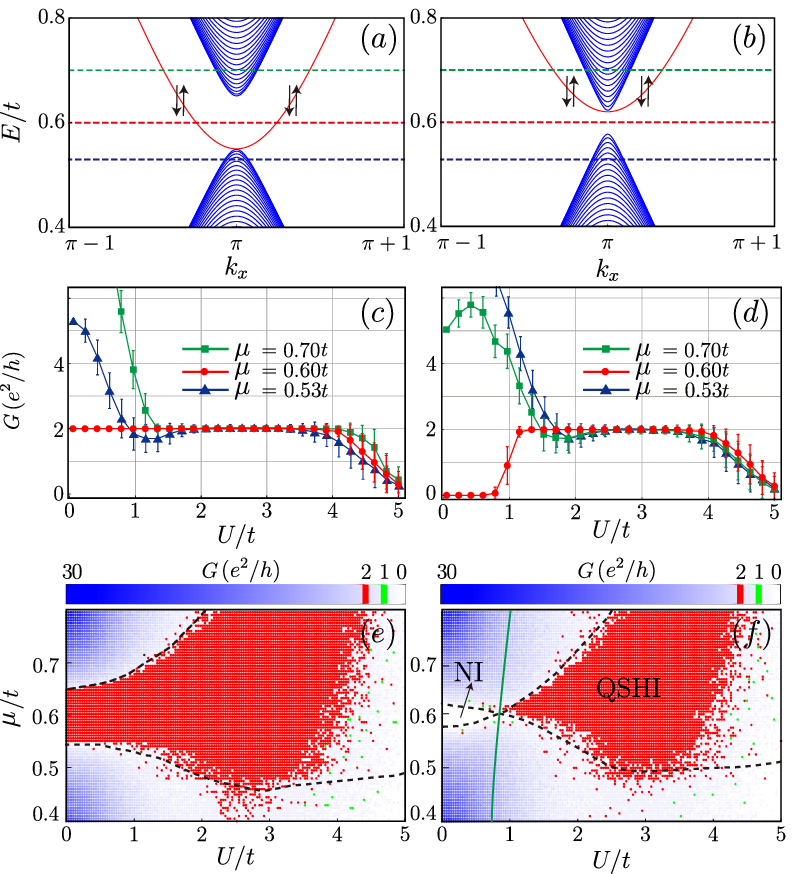}\caption{(Color online) (a) Energy spectrum of the clean system with the open boundary condition along the $y$ direction and $S_\pm=0.55t$. The blue curves are the bulk subbands and the red lines correspond to the edge states. The up and down arrows denote the spin orientation. (c) The conductance as a function of the strength of disorder $U$ at different chemical potentials $\mu=0.70t$, $0.60t$ and $0.53t$, which correspond to the green, red and blue dashed lines shown in (a). The error bars show the standard deviation of the conductance for $1000$ samples. (e) Phase diagram for the conductance as functions of the disorder strength $U$ and the chemical potential
$\mu$. The green solid and black dashed lines obtained from the self-consistent Born approximation denote the phase boundary defined as $\Delta_{\tilde{E}}=0$ and $\tilde{E}_2 < \tilde{\mu}< \tilde{E}_1$. The other panels (b), (d) and (f) are the same as (a), (c) and (e), except that $S_\pm=0.62t$. In our numerical simulations, the width and length of the sample are chosen to be $L_y =200a$ and $L_x = 300a$.
}%
\label{fig2}%
\end{figure}

We first calculate the conductance $G$ as a function of disorder in the case of no spin-dependent staggered potential (i.e., $\Delta_{s}=0$, $S_{+}=S_{-}=\Delta_{c}\neq 0$). In this case, the time-reversal symmetry is respected and the upper gap is topologically non-trivial when $S_\pm<2\lambda$. To better understand the different phases in the clean system, we plot the energy spectrum of a strip of sample for different staggered potential strengths $S_\pm=0.55t$ (Fig. \ref{fig2} (a)) and $S_\pm=0.62t$ (Fig. \ref{fig2} (b)), which correspond to the QSHI state and NI state respectively. As shown in Fig. \ref{fig2}(a), there
exists a pair of helical edge states connecting the conduction and valence band since in this case $S_\pm<2\lambda$. While for $S_\pm=0.62t>2\lambda$ (Fig. \ref{fig2}(b)), the helical edge states merge into the bulk states, resulting in a NI state.

Figs. \ref{fig2}(c-d) show the numerical results of the longitudinal conductance $G$ versus the disorder strength $U$ at three different chemical potentials $\mu=0.70t$, $0.60t$ and $0.53t$. When the system is in the QSHI regime ($S_\pm=0.55t$) and the chemical potential located in the band gap ($\mu=0.60t$), the conductance is quantized to be $2e^2/h$ and quite stable as long as the disorder strength $U$ is smaller than $U_c=4t$. This is expected since the conductance is attributed to the topologically helical edge states. Interesting phenomena emerges in the NI regime ($S_\pm=0.62t$) when we still keep the chemical potential in the band gap. Initially, the conductance is zero since there is no electronic state inside the band gap for a NI. With the increase of disorder strength, the conductance rises and reaches to a quantized value of $2e^{2}/h$, then a plateau is formed before $G$ decreases and eventually vanishes. For the chemical potential lies in the conduction and valence bands ($\mu=0.70t$ and $\mu=0.53t$), the conductance is suppressed by disorder and then also forms a quantized plateau before it continues to drop.

The absence of fluctuation in the disorder-induced quantized conductance plateau of $2e^{2}/h$ in Figs. \ref{fig2}(c) and \ref{fig2}(d) suggests that it is attributed to the helical edge states of QSHI. This disorder-induced QSHI phase is nothing else but the TAI phase discovered a few years ago.

Figures \ref{fig2}(e) and (f) show the phase diagrams for initial states are the QSHI state and the NI state respectively. Each point corresponds to a single realization of the disorder potential, which turns out to be sufficient for determining the region of the disorder induced QSHI phase. We can see that there exits a large phase area of QSHI in the $(U,\mu)$ phase space where the conductance is quantized.

\subsection{Coexisting of spin-independent and dependent staggered potentials}

Now we proceed to study transport properties in the presence of both the spin-independent and dependent staggered potentials. The spin-dependent potential breaks the time-reversal symmetry and can result in the QAHI state or NI state for the upper gap depending on the competition between staggered potentials and ISOC.

\begin{figure}[ptb]
\includegraphics[width=8cm]{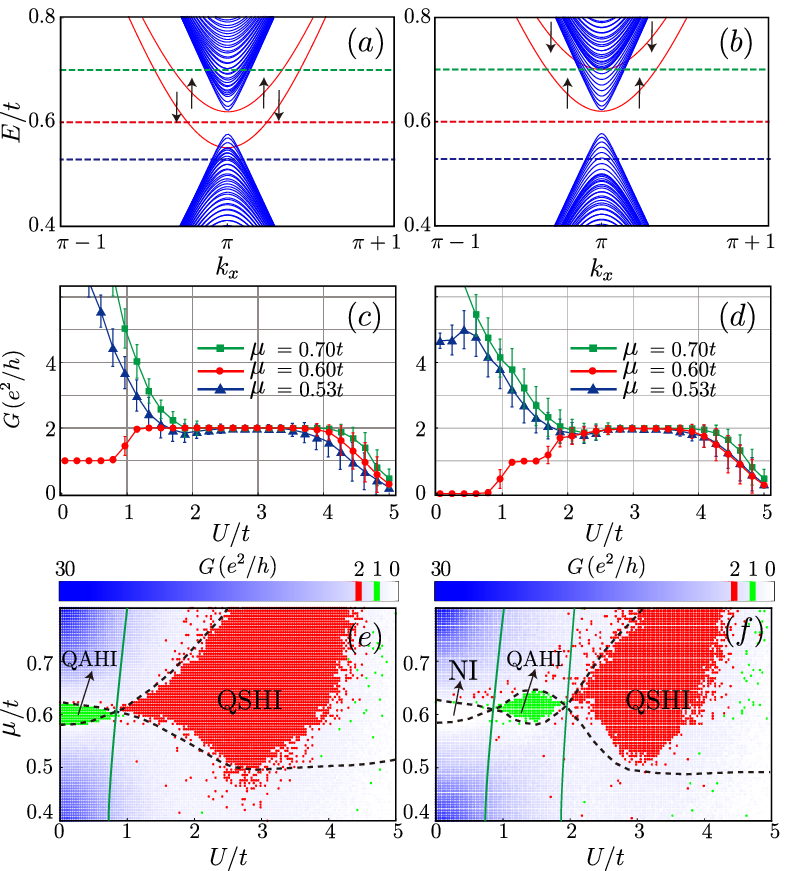}\caption{(Color online) (a) Energy spectrum of the clean system with the open boundary condition along the $y$ direction for $S_+=0.62t$ and $S_-=0.55t$. The blue curves are the bulk subbands and the red lines correspond to the edge states. The up and down arrows denote the spin orientation. (c) The conductance as a function of the strength of disorder $U$ at the chemical potentials $\mu=0.70t$, $0.60t$ and $0.53t$, which are labeled with the green, red and blue dashed lines shown in (a). The error bars show the standard deviation of the conductance for $1000$ samples. (e) Phase diagram for the conductance as functions of the disorder strength $U$ and the chemical potential $\mu$. The green solid and black dashed lines obtained from the self-consistent Born approximation correspond to the phase boundary defined as $\Delta_{\tilde{E}}=0$ and $\tilde{E}_2 < \tilde{E}_f < \tilde{E}_1$. The other panels (b), (d) and (f) are the same as (a), (c) and (e), except that $S_+=0.62t$ and $S_-=0.70t$. In our numerical simulations, the width and length of the strip Lieb model are chosen to be $L_y =200a$ and $L_x = 300a$.}%
\label{fig3}%
\end{figure}

The band structures of pure sample are shown in Figs. \ref{fig3}(a) and (b). For $S_+=0.62t$ and $S_-=0.55t$, a single topologically chiral edge state appears in the band gap, indicating the QAHI state. In the case of $S_\pm>2\lambda$ with $S_+=0.62t$ and $S_-=0.70t$, there is no edge state in the gap, corresponding to the NI state.

We compute the conductance $G$ as a function of the strength of disorder for the chemical potential is located in the conduction band ($\mu=0.70t$), valence band ($\mu=0.53t$) and band gap ($\mu=0.60t$). The results are shown in Figs. \ref{fig3}(c-d). For chemical potential located in the conduction and valence band, the results are quite similar to the case with only spin-independent staggered potential. As increasing the strength of disorder, the conductance $G$ is suppressed and then forms a quantized conductance plateau of $2e^2/h$, finally disappears. It means that we have a disorder-induced QSHI state even when the time-reversal symmetry is broken. Even more interesting is the case with the chemical potential located in the band gap. $G$ is quantized to be $e^2/h$ if we start with the QAHI state. Surprisedly, with the increase of the strength of disorder, $G$ goes through a transition from the quantized plateau of $e^2/h$ to $2e^2/h$, suggesting a disorder-induced phase transition from the QAHI state to the time-reversal-symmetry-broken QSHI state. If we start with the NI state and increase the disorder, first, the conductance plateau of $e^2/h$ is formed, then subsequently the conductance plateau of $2e^2/h$ emerges, suggesting a succession of topological phase transitions, including a NI-QAHI phase transition and a successive QAHI-QSHI phase transition. We can see that the QSHI state is favored in the time-reversal-symmetry-broken system when the disorder is strong enough. The reason is that the disorder renormalizes the staggered potentials and $S_\pm$ are effectively decrease as shown in Eq. (\ref{RSP}).

The phase diagrams in the $U$-$\mu$ parameter space obtained by numerical simulations are shown in Figs. \ref{fig3}(e-f). The latter phase diagram shows the QAHI phase at intermediate disorder and the time-reversal-symmetry-broken QSHI phase at stronger disorder.

We would like to emphasize that it is different from the time-reversal-invariant case and previous results based on the BHZ model, the disorder can induce both the QAHI state and the time-reversal-symmetry-broken QSHI state in the Lieb lattice model with the spin-dependent staggered potential. Of particular interest is the disorder-tunable switching between QAHI phase and QSHI phase.

\subsection{Nonequilibrium local current distribution}
\begin{figure}[ptb]
	\includegraphics[width=8cm]{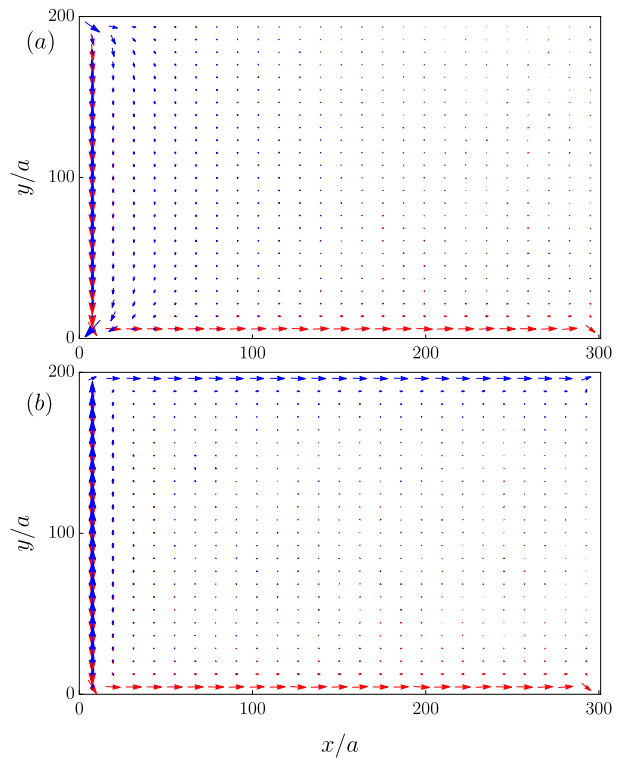}\caption{(Color online) The averaged nonequilibrium local current distribution for the disorder-induced topological phases of the Lieb lattice model with $S_+=0.62t$, $S_-=0.70t$ and $\mu=0.60t$. We take disorder strength $U=1.5t$ and $U=3t$ in (a) and (b) respectively. The red and blue arrows correspond to the spin-up and spin-down components of the local current, respectively. The arrow size means the strength of the local current.
}%
\label{fig4}%
\end{figure}
To further confirm the assertion that the quantized conductance plateaus
originate from the robust edge states, we calculate the nonequilibrium local
current distribution between neighboring sites $\mathbf{{i}}$ and
$\mathbf{{j}}$ from the following formula\cite{Jiang09PRB}
\begin{equation}
J_{\mathbf{{i}\rightarrow{j}}}=\frac{2e^{2}}{h}\text{Im} \left[  \sum
_{\alpha,\beta}{H_{\mathbf{{i} \alpha,{j}\beta}}G^{n}_{\mathbf{{j}\beta
			,{i}\alpha}}}\right]  \left(  V_{L}-V_{R} \right)  \text{,}%
\end{equation}
where $V_{L}(V_R)$ describes the voltage of the left (right) lead, and
$G^{n}=G^{r}\Gamma_{L} G^{a}$ is the electron correlation function. To calculate the local current distribution, a small external bias $V=V_L-V_R$ is applied longitudinally between the two terminals, where $V_L$ and $V_R$ describe the voltages of the left and right leads. The small bias voltage $V$ is fixed to be $0.0001t/e$. We assume the electrostatic potential in the central part is $\phi(i_x)=(L_x-i_x+1)V/(L_x+1)$, where $i_x$ is the site index along the $x$-direction and $1\leq i_x \leq L_x$. Then, the electric field is uniformly distributed in the central sample region.

In Figs. \ref{fig4}(a-b), we show the current distribution of the disorder-induced QAHI state and disorder-induced QSHI state. The red and blue arrows correspond to the local currents of the spin-up and down species, respectively. For the disorder-induced QAHI phase with $G=e^2/h$, only spin-up edge current propagates along the lower boundary of the sample since the QAHI state comes from the spin-up species. While for the disorder-induced QSHI state, the spin-up and spin-down edge currents are localized at the two opposite sides of the sample respectively.

\section{Self-consistent Born approximation}

To corroborate the numerical simulations, we analyze the present model
within an effective medium theory based on the Born approximation in which
high order scattering processes are neglected \cite{Groth09PRL}. In the
self-consistent Born approximation, the self-energy $\Sigma$ caused by disorder is given by the following integral equation
\begin{equation}
\Sigma=\frac{U^{2}}{12}(\frac{a}{2\pi})^{2}\int_{FBZ}d\mathbf{k}%
\frac{1}{\mu-H\left(  \mathbf{k}\right)  -\Sigma}\text{.}
\label{Anderson}%
\end{equation}
The coefficient 1/12 comes from the variance $\left\langle U^{2}\right\rangle =U^{2}/12$ of a random variable uniformly distributed in
the range $\left[-U/2,U/2\right]$. This integration is over the first Brillouin zone (FBZ).

After some calculations, we found that disorder has a renormalized
effect on the model parameters, leading to phase transitions on Lieb lattices. By calculating the conduction band minimum $\tilde{E}_{1}$ and the valence band maximum $\tilde{E}_{2}$ of the renormalized Hamiltonian ($\tilde{H}=H+\Sigma$) as functions of $\mu$ and $U$, we obtain the curves in Figs. \ref{fig2}(e-f) and Figs. \ref{fig3}(e-f). The green curves denote the phase boundary line $\tilde{E}%
_{1}-\tilde{E}_{2}=\Delta_{\tilde{E}}=0$, which means the gap closing.
The region between black dashed curves is obtained by $\tilde
{E}_{2}<\tilde{\mu}<\tilde{E}_{1}$. It is evident that the analytical results fit well with the numerical results.

After a low-energy expansion of $H(\mathbf{k})$ at $\mathbf{k}=(\pi,\pi)$ and setting $\Sigma=0$ on the right-hand side of Eq. (\ref{Anderson}), the integral can be calculated analytically with \cite{Groth09PRL,GUO10PRL}
\begin{equation}\label{RSP}
\tilde{S}_{\pm }=S_{\pm }-\frac{\pi U^{2}}{192\left( \mu+S_{\pm }\right)},
\end{equation}%
where we keep only the leading ultraviolet-divergent part of the integral and $\tilde{S}_{\pm}=\tilde{\Delta}_c-\tilde{\Delta}_s$ represents the strength of the renormalized staggered potential. It is found that the disorder diminishes $S_{\pm}$. From Eq. (\ref{RSP}) we can conclude that the phase transition occurs at $U_c^\pm=\sqrt{192(\mu+S_\pm)(S_\pm-2\lambda)/\pi}$.

\section{Conclusion}

In this paper, we investigate the disorder-induced topological phase transitions on spin-orbit coupled Lieb lattices with spin-independent and dependent potentials. We have observed distinct phase transitions by numerical simulations. When only the spin-independent staggered potential exists, we found a disorder-induced QSHI phase with time-reversal symmetry. When two kinds of staggered potential coexist, Anderson disorder can induce the QAHI phase and the time-reversal-symmetry-broken QSHI phase as well as switch two different topological phases. We map out the phase diagrams in the parameter space. To confirm the numerical results, we use the effective medium theory based on the self-consistent Born approximation. The analytical results agree very well with our numerical results.

Finally, we would like to point out the electron interaction may play an important role in Lieb lattices due to the flat band. Some novel topological phases due to the electron interaction have been found in this system like the topological Mott insulator \cite{Dauphin16PRA}. Although in this paper we concentrate on the non-interacting Lieb lattice system, we believe that the interplay of disorder, topology and interaction \cite{Hung16PRB} on Lieb lattices would produce more new exotic quantum phenomena.

\section*{Acknowledgments}

B. Zhou was supported by the National Natural Science Foundation of China (Grant No. 11274102), the Program for New Century Excellent Talents in University of Ministry of Education of China (Grant No. NCET-11-0960), and the Specialized Research Fund for the Doctoral Program of Higher Education of China (Grant No. 20134208110001). R. Chen and D.-H. Xu are supported by the National Natural Science Foundation of China (Grant No. 11704106)

\end{document}